\documentclass{ws-procs9x6}

\def\de{\delta}

\def\et{\eta}
\def\th{\theta}

\def\ta{\tau}

\def\fr#1#2{{{#1} \over {#2}}}

\def\frac#1#2{{\textstyle{{#1}\over {#2}}}}

\def\lsim{\mathrel{\rlap{\lower4pt\hbox{\hskip1pt$\sim$}}
    \raise1pt\hbox{$<$}}}
\def\gsim{\mathrel{\rlap{\lower4pt\hbox{\hskip1pt$\sim$}}
    \raise1pt\hbox{$>$}}}
\def\sqr#1#2{{\vcenter{\vbox{\hrule height.#2pt
         \hbox{\vrule width.#2pt height#1pt \kern#1pt
         \vrule width.#2pt}
         \hrule height.#2pt}}}}

\def\lrpartial{\raise 1pt\hbox{$\stackrel\leftrightarrow\partial$}}

\def\etal{{\it et al.}}

\newcommand{\beq}{\begin{equation}}
\newcommand{\eeq}{\end{equation}}
\newcommand{\bea}{\begin{eqnarray}}
\newcommand{\eea}{\end{eqnarray}}
\newcommand{\rf}[1]{(\ref{#1})}

\begin{document}

\title{Lorentz and CPT tests involving antiprotons}

\author{RALF LEHNERT}

\address{Department of Physics and Astronomy\\
Vanderbilt University,
Nashville, Tennessee, 37235\\ 
E-mail: ralf.lehnert@vanderbilt.edu}

\maketitle

\abstracts{
Perhaps the largest gap 
in our understanding of nature at the smallest scales 
is a consistent quantum theory 
underlying the Standard Model and General Relativity. 
Substantial theoretical research has been performed in this context, 
but observational efforts are hampered 
by the expected Planck suppression of deviations from conventional physics. 
However, a variety of candidate models 
predict minute violations of both Lorentz and CPT invariance. 
Such effects open a promising avenue 
for experimental research in this field 
because these symmetries are amenable to Planck-precision tests.\\
\vskip-8pt
\hspace{.5cm} The low-energy signatures of Lorentz and CPT breaking 
are described 
by an effective field theory called the Standard-Model Extension (SME). 
In addition to the body of established physics 
(i.e., the Standard Model and General Relativity), 
this framework incorporates all Lorentz- and CPT-violating corrections 
compatible with key principles of physics. 
To date, the SME has provided the basis 
for the analysis of numerous tests of Lorentz and CPT symmetry 
involving protons, neutrons, electrons, muons, and photons. 
Discovery potential exists in neutrino physics.\\
\vskip-8pt
\hspace{.5cm} A particularly promising class of Planck-scale tests 
involve matter--antimatter comparisons at low temperatures. 
SME predictions for transition frequencies in such systems 
include both matter--antimatter differences and sidereal variations. 
For example, in hydrogen--antihydrogen spectroscopy, 
leading-order effects in a 1S--2S transition 
as well as in a 1S Zeeman transition could exist 
that can be employed to obtain clean constraints. 
Similarly, 
tight bounds can be obtained 
from Penning-trap experiments involving antiprotons.
}

\newpage

\section{Motivations}
\label{mot}

The study of the hydrogen (H) atom
is closely associated with
two of the most important achievements in 20th-century physics:\cite{H}
the explanation of its discrete spectrum
provided a great triumph for quantum theory,
and many details of the H spectrum
serve as a powerful testimonial for Special Relativity
in the microscopic world of atoms.
The combination of these two groundbreaking theories
lies at the foundation
of the most successful physical theory 
to date---the Standard Model of particle physics. 

For roughly half a century,
theoretical research in fundamental physics 
has been dominated
by various approaches to synthesize
the Standard Model and General Relativity 
into a single unified theory
that incorporates, for example,
a quantum description of the gravitational interaction.
Although a substantial amount of progress
has been achieved on the theoretical front,
the expected Planck suppression of quantum-gravity effects 
hampers experimental research in this field:\cite{pop}
low-energy measurements are likely to require sensitivities
of at least one part in $10^{17}$.
This talk argues
that the recent creation and observation 
of hot antihydrogen ($\overline{\textrm{H}}$),\cite{hbar} 
the subsequent production of cold $\overline{\textrm{H}}$
by the ATHENA and ATRAP collaborations,\cite{cold}
and the synthesis of antiprotonic Helium 
by the ASACUSA collaboration\cite{asa}
pave the way for tests
that could shed some light on this issue.
The basic ideas behind this belief are summarized
in the remainder of this section.

The presumed minuscule size of candidate quantum-gravity signatures
requires a careful choice of experiments.
A promising avenue 
that one can pursue
is testing physical laws 
that satisfy three primary criteria.
First,
one should consider presently established fundamental laws
that are believed to hold {\it exactly}.
Any measured deviations
would then definitely indicate qualitatively new physics.
Second,
the chances of finding an effect are increased
by testing laws
that can be {\it violated}
in credible candidate fundamental theories.
Third,
from a practical viewpoint,
this law must be amenable to {\it ultrahigh-precision} tests.

An example of a physics law
that satisfies all of these criteria
is CPT invariance.
As a brief reminder,
this law requires 
that the physics remains invariant 
under the combined operations
of charge conjugation (C), 
parity inversion (P),
and time reversal (T).
Here, 
the C transformation links particles 
and antiparticles,
P corresponds to a spatial reflection 
of physics quantities
through the coordinate origin, 
and T reverses a given physical process in time.\cite{cpt}
The Standard Model
is CPT-invariant by construction,
so that the first criterion is satisfied.
With regards to criterion two, 
we mention 
that a variety of approaches to fundamental physics
can lead to CPT breakdown.
Examples include
strings,\cite{kps}
spacetime foam,\cite{sf}
nontrivial spacetime topology,\cite{klink}
and cosmologically varying scalars.\cite{varscal}
The third criterion is met as well.
Consider,
for instance,
the conventional figure of merit for CPT conservation 
in the Kaon system:
its value lies currently at $10^{-18}$, 
as quoted by the Particle Data Group.\cite{pdg}

The CPT transformation
relates a particle to its antiparticle.
One would therefore expect
that CPT invariance implies a symmetry 
between  matter and antimatter.
Indeed,
one can prove
that the magnitude of the mass, charge, decay rate, gyromagnetic ratio,
and other intrinsic properties of a particle
are exactly equal to those of its antiparticle.
This prove can be extended
to systems of particles and their dynamics.
For example,
atoms and anti-atoms
must exhibit identical spectra
and a particle-reaction process
and its CPT-conjugate process 
must possess the same reaction cross section.
It follows
that experimental matter--antimatter comparisons
can serve as probes for the validity of CPT symmetry.
In particular,
the extraordinary sensitivities 
offered by atomic spectroscopy 
suggest comparative tests
with H and $\overline{\textrm{H}}$
as high-precision tools
in this context.

CPT violation in Nature would also lead to 
other, less obvious effects. 
The celebrated CPT theorem of Bell, L\"uders, and Pauli states
that CPT invariance arises under a few mild assumptions
through the combination of 
quantum theory and Lorentz symmetry. 
If CPT invariance is broken
one or more of the assumptions 
necessary prove the CPT theorem
must be false.
This leads to the obvious question
which one of the fundamental assumptions in the CPT theorem
has to be dropped.
Since both CPT and Lorentz symmetry
involve spacetime transformations,
it is natural to suspect
that CPT violation implies Lorentz-invariance breakdown.
This has recently been confirmed rigorously
in Greenberg's ``anti-CPT theorem,''
which  roughly states
that in any local, unitary, relativistic point-particle field theory
CPT violation implies Lorentz violation.\cite{green02,green}
Note, however, 
that the converse of this statement---namely 
that Lorentz violation implies CPT breaking---is not true in general.
{\it In any case,
it follows that CPT tests also probe Lorentz symmetry.}
This result offers the possibility for 
a another class of CPT-violation searches
in addition to instantaneous matter--antimatter comparisons:
probing for sidereal effects
in matter--antimatter and other systems.

This talk gives an overview of Lorentz and CPT violation
as a tool in the search for underlying physics---possibly 
arising at the Planck scale.
Section \ref{lv} discusses some forms of Lorentz and CPT violation
that could be considered when constructing test models.
The Standard-Model Extension (SME),
which is currently the standard and most general framework
for CPT and Lorentz tests,
is reviewed in Sec.\ \ref{smesec}.
Section \ref{mech} gives two examples 
of how a Lorentz- and CPT-invariant model 
can lead to the violation of these symmetries in the ground state
generating SME coefficients.
In Sec.~\ref{test},
some experimental tests of Lorentz and CPT invariance
that involve antimatter
are discussed.
Section \ref{summ}
contains a brief summary.

\section{Types of Lorentz violation}
\label{lv}

The first step in constructing a test model
parametrizing the breakdown of Lorentz and CPT symmetry is 
to determine possible types of Lorentz violation.
Additional considerations for CPT breakdown
are unnecessary by the anti-CPT theorem
because general Lorentz violation
will include CPT breaking.
It turns out
that a clear understanding of the 
{\it fundamental principle of coordinate independence}
will provide us 
with a useful, rough classification of types of Lorentz violation.
For this reason,
it appears appropriate begin with a brief review of this fundamental principle
and its implementation. 

Coordinate independence is one of the most basic principles in physics. 
Its need in the presence of Lorentz breaking 
is well established,
and it has served as the basis 
for the construction of the SME.\cite{sme,thres}  
However,
this principle is sometimes not fully appreciated.
For example,
in some investigations of Lorentz and CPT violation 
coordinate-{\it dependent} physics emerges, 
and occasionally Lorentz-symmetry breakdown is identified 
with the loss of coordinate independence. 

A given labeling scheme for events in space and time 
is called a coordinate system. 
Such a labeling 
typically depends on the observer choosing the coordinates,
and it is thus arbitrary to a large extent. 
In other words,
coordinate systems are {\it mathematical tools}
for the measurement, description, and prediction
of physical phenomena.
But since they are a pure product of human thought,
coordinates {\it lack physical reality}.
It follows that the physics should remain
unaffected by the choice of a particular coordinate system.
This principle of coordinate independence 
is one of the most fundamental in science. 
Since it assures 
that the physics remains independent of the observer,
it is also called {\it observer invariance}. 
It should therefore be possible
to formulate the fundamental laws of physics
in a coordinate-free language.
For example,
this can be achieved mathematically,
when spacetime is given a manifold structure 
and physical quantities are represented 
by geometric objects, 
such as tensors or spinors. 

The principle of coordinate invariance is 
more fundamental than Lorentz symmetry. 
Consider,
for instance, 
Newton's second law of motion in nonrelativistic classical mechanics
\beq
\label{second}
\vec{F}=m\dot{\vec{v}}\,,
\eeq 
as well as its relativistic version
\beq
\label{relsecond}
F^{\mu}=m\:\fr{dv^{\mu}}{d\ta}\,,
\eeq 
where $\tau$ denotes the mass' proper time,
and $F^{\mu}$ is the usual relativistic generalization of the force $\vec{F}$.
Both laws are coordinate independent.
Equation \rf{second} takes the same form 
in all inertial gallilean frames;
it is formulated in the coordinate-free language of 3-vectors.
Similarly, Eq.\ \rf{relsecond}
remains of the same form in all inertial Minkowski coordinate systems;
it is expressed in terms of 4-vectors.
However,
only Eq.\ \rf{relsecond} is Lorentz invariant.
We conclude 
that coordinate independence is more general than Lorentz symmetry
because there might very well be laws---such as Eq.\ \rf{second}---that 
maintain coordinate invariance
but violate Lorentz symmetry.

The above situation becomes even more transparent
with the following observation.
Lorentz symmetry does not only require coordinate independence,
but it also dictates the transformations
that relate different inertial frames.
Although Eq.\ \rf{second}
is coordinate independent,
inertial frames are related by Gallilei
instead of Lorentz transformations.
Mathematically speaking,
both cases allow a coordinate-invariant spacetime-manifold description,
but the manifold structure is gallilean in the case of Eq.\ \rf{second}
and lorentzian in the case of Eq.\ \rf{relsecond}.
The question which spacetime manifold is realized in Nature
must be answered experimentally.

{\bf Lorentz violation through non-lorentzian manifolds.} 
The above considerations lead to one possible type of Lorentz violation
maintaining coordinate independence:
the spacetime manifold could be non-lorentzian. 
Then,
the fundamental physics laws
have the same form in all inertial frames,
but the frames are no longer related by the usual Lorentz transformations.
This point of view is taken 
in the early relativity test model of Robertson 
and its extension by Mansouri and Sexl,
as well as in the so called ``doubly special relativities.''
In the present work,
we do not treat this type of Lorentz-symmetry violation separately
because it is known that (at least some of) these effects
are equivalent to those of another type of Lorentz violation
discussed next.
Moreover,
such frameworks are typically purely kinematical
precluding the analysis of atomic level shifts,
for example.

{\bf Lorentz violation through a nontrivial vacuum.} 
Most modern approaches to fundamental physics
involve lorentzian manifolds,
where inertial frames are related by the usual Lorentz transformations.
Such approaches take Lorentz symmetry 
as a key ingredient,
and the issue arises
as to whether Lorentz violation can occur in such situations.
It turns out 
that this is indeed the case
if the vacuum contains a tensorial background.
The primary emphasis in this section
is to gain some intuitive understanding
of Lorentz breakdown in the presence of such a background.
The question of how to generate tensorial backgrounds
in a Lorentz-invariant theory
is deferred to Sec.\ \ref{mech}.

\begin{figure}
\begin{center}
\includegraphics[width=0.95\hsize]{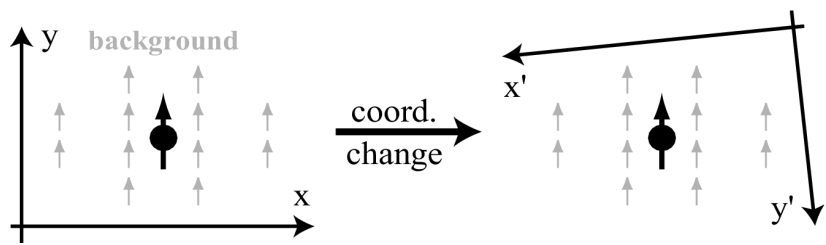}
\end{center}
\caption{Coordinate independence. 
Two experimenters
observe identical physical systems 
represented by the black ``particle with spin.''
Although they may choose to employ different coordinate systems
to describe their observations,
the outcome of the experiment remains unaffected 
by this choice.
It must therefore be possible
to relate observations
recorded with respect to different reference frames
by appropriate transformations of coordinates.
The principle of coordinate independence
therefore assures
that the physics is independent of the observer.
}
\label{fig1} 
\end{figure}

\begin{figure}
\begin{center}
\includegraphics[width=0.95\hsize]{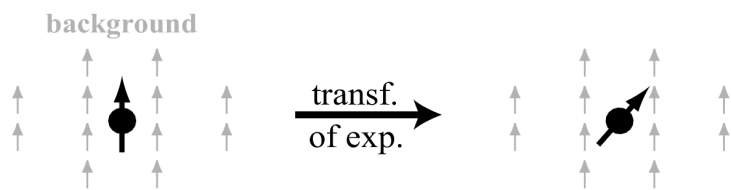}
\end{center}
\caption{Particle transformations.
Tests of rotational invariance, for example,
would {\it not} be carried out as in Fig.\ \ref{fig1}:
identical experiments with the observer rotated.
Instead,
one would perform a suitable measurement,
repeat it with a rotated apparatus,
and then compare the two measurements.
Under these types of transformations,
which involve localized particles and fields
and leave unchanged the background,
symmetry can be lost
because of the different orientation
with respect to the vacuum structure.
}
\label{fig2} 
\end{figure}

It is again useful
to consider a familiar example from classical physics.
Suppose the particle described by Eq.\ \rf{relsecond}
has charge $q$
and is subjected to an {\it external} electromagnetic field $F^{\mu\nu}$.
We remind the reader 
that the components of $F^{\mu\nu}$
are determined 
by the usual electric and magnetic fields $\vec{E}$ and $\vec{B}$.
The left-hand side of Eq.\ \rf{relsecond}
is now given by the Lorentz force,
which reads $qF^{\mu\nu}v_{\nu}$ in covariant form.
The equation of motion 
determining the trajectory of our particle
becomes
\beq 
qF^{\mu\nu}v_{\nu}=m\:\fr{dv^{\mu}}{d\ta}\; . 
\label{example} 
\eeq 
Note that Eq.\ \rf{example} remains valid 
in {\it all} inertial coordinate systems 
because it is a tensor equation. 
Invariance under Lorentz transformations of the coordinate system
is therefore maintained. 
However, 
the external $F^{\mu\nu}$ background breaks, 
for example, 
symmetry under arbitrary rotations of the charge's trajectory. 
Among the consequences of this rotation-symmetry violation 
is the non-conservation of the particle's angular momentum.
Notice the difference to coordinate changes, 
which leave unaffected the physics: 
in the present case, 
only the trajectory is rotated, 
so that its orientation 
with respect to $F^{\mu\nu}$ 
can change. 
One then says that particle Lorentz symmetry is violated, 
despite the presence of Lorentz coordinate independence.\cite{sme,thres} 
Figures \ref{fig1} and \ref{fig2}
illustrate the difference
between particle Lorentz transformations
and Lorentz coordinate transformations.

Although the above example captures the main features
of Lorentz violation through background vectors or tensors,
there is an important difference to situations 
where these vectors or tensors 
can be considered as part of the effective vacuum.
Our above background $F^{\mu\nu}$ is a local electromagnetic field 
caused by other 4-currents 
that can in principle be accessed experimentally. 
Such backgrounds are therefore {\it not} a feature of the vacuum,
so that these situations cannot be considered 
as exhibiting fundamental Lorentz violation. 
This is to be contrasted 
with situations involving candidate underlying physics,
where tensorial backgrounds can extend over the entire Universe
and are outside of experimental control.
Such backgrounds {\it must} be viewed 
as a property of the effective vacuum,
which can then be considered 
as violating Lorentz symmetry (see Sec\ \ref{mech}).

As a result of the lorentzian structure of the underlying manifold 
and the usual Lorentz-covariant dynamics at the fundamental level, 
this approach appears closest to established theories. 
The physical effects in such Lorentz-breaking vacua 
are perhaps comparable to those inside certain crystals: 
the physics remains independent of the chosen coordinate system, 
but particle propagation, 
for example, 
can depend upon the direction. 
An immediate consequence is
that one can locally still work 
with the metric $\et^{\mu\nu}={\rm diag}(1,-1,-1,-1)$, 
particle 4-momenta are still additive 
and still transform in the usual way under coordinate changes, 
and the conventional tensors and spinors 
still represent physical quantities.

{\bf Models with coordinate-dependent physics.} 
We have argued above
that coordinate independence
is a principle more fundamental than Lorentz symmetry.
If one is willing to give up 
coordinate independence,
the loss of Lorentz invariance is unsurprising. 
Although it seems to be impossible 
to perform meaningful scientific investigations 
involving coordinate-{\it dependent} physics, 
such approaches to Lorentz breaking 
have been considered in the literature. 
More specifically, 
there have been the following two attempts 
in the context of neutrino phenomenology: 
the first one forces the masses of particle and antiparticle 
to be different,\cite{baren01} 
while the second one suggests to construct a model 
from positive-energy eigenspinors only.\cite{baren02} 
Both of these approaches are known 
to involve coordinate-dependent off-shell physics.\cite{green02,green03}
In what follows, 
we do not consider these models further.

\section{The Standard-Model Extension}
\label{smesec}

The next step after determining general low-energy manifestations
of Lorentz violation 
is the identification 
of specific experimental signatures for these effects 
and the theoretical analysis of Lorentz-violation searches. 
This task is most conveniently accomplished 
employing a suitable test model. 
Many Lorentz tests are motivated and analyzed 
in purely kinematical frameworks 
allowing for small violations of Lorentz symmetry. 
Examples are  the aforementioned Robertson's framework 
and its Mansouri--Sexl extension, 
as well as the $c^2$ model 
and phenomenologically constructed modified dispersion relations. 
However, 
the implementation of general dynamical features 
significantly increases the scope 
of Lorentz tests. 
For this reason, 
the SME 
mentioned in the Sec.\ \ref{mot} 
has been developed. 
But the use of dynamics in Lorentz-violation searches 
has recently been questioned 
on grounds of test-framework dependence. 
We disagree with this assertion  
and begin with a few arguments in favor of a dynamical test model. 

The construction of a dynamical test framework 
is constrained by the requirement 
that known physics must be recovered 
in certain limits, 
despite some residual freedom 
in introducing dynamical features 
compatible with a given set of kinematical rules. 
Moreover, 
it appears difficult 
and may even be impossible 
to develop an effective theory containing the Standard Model 
with dynamics significantly different from that of the SME. 
We also point out
that kinematical analyses 
are limited 
to only a subset of potential Lorentz-violating signatures 
from fundamental physics. 
From this viewpoint, 
it seems to be desirable 
to explicitly implement dynamical features 
of sufficient generality 
into test models for Lorentz and CPT symmetry.  

{\bf The generality of the SME.}
In order to understand 
the generality of the SME, 
we review the main cornerstones of its construction.\cite{sme} 
Starting from the usual Standard-Model lagrangian ${\mathcal L}_{\rm SM}$, 
Lorentz-violating modifications $\de {\mathcal L}$ are added: 
\beq
{\mathcal L}_{\rm SME}={\mathcal L}_{\rm SM}+\de {\mathcal L}\; ,
\label{sme}
\eeq
where the SME lagrangian is denoted by ${\mathcal L}_{\rm SME}$. 
The correction term $\de {\mathcal L}$ 
is obtained by contracting Standard-Model field operators 
of any dimensionality 
with Lorentz-violating tensorial coefficients 
that describe the nontrivial vacuum 
discussed in the previous section. 
To guarantee coordinate independence, 
this contraction must give 
observer Lorentz scalars. 
It becomes thus apparent 
that all possible contributions to $\de {\mathcal L}$ 
yield the most general effective dynamical description 
of Lorentz violation 
at the level of observer Lorentz-invariant quantum field theory. 
For simplicity,
we have focused on nongravitational physics 
in the above construction.
We mention
that the complete SME 
also contains an extended gravity sector. 

Possible Planck-scale features, 
such as non-pointlike elementary particles 
or a discrete spacetime, 
are unlikely to invalidate 
the above effective-field-theory approach 
at currently attainable energies. 
On the contrary, 
the phenomenologically successful Standard Model 
is widely believed
to be an effective-field-theory approximation 
for underlying physics. 
If fundamental physics 
indeed leads to minute Lorentz-breaking effects, 
it would seem contrived 
to consider low-energy effective models 
outside the framework of effective quantum field theory. 
We finally note 
that the necessity for a low-energy description 
beyond effective field theory 
is also unlikely to arise 
in the context of candidate fundamental models 
with novel Lorentz-{\it invariant} aspects, 
such as additional particles, 
new symmetries, 
or large extra dimensions. 
Lorentz-symmetric modifications 
can therefore be implemented into the SME, 
if needed.\cite{susy} 

{\bf Advantages of the SME.}
The SME 
allows the identification 
and direct comparison 
of virtually all currently feasible experiments
searching for Lorentz and CPT breaking. 
Moreover, 
certain limits of the SME 
correspond to classical kinematics test models of relativity 
(such as the previously mentioned Robertson's framework, 
its Mansouri-Sexl extension, 
or the $c^2$ model).\cite{km02} 
Another advantage of the SME 
is the possibility of implementing 
further desirable conditions 
besides coordinate independence. 
For example, 
one can choose to impose 
spacetime-translation invariance, 
SU(3)$\times$SU(2)$\times$U(1) gauge symmetry, 
power-counting renormalizability, 
hermiticity,
and pointlike interactions. 
These requirements 
further restrict the parameter space for Lorentz violation. 
One could also adopt simplifying choices, 
such as a residual rotation symmetry
in certain coordinate systems. 
This latter assumption 
together with additional simplifications of the SME 
has been considered in Ref.\ \refcite{cg99}. 

{\bf Analyses performed within the SME.}
To date, 
the flat-spacetime limit 
of the minimal SME
has provided the basis 
for numerous experimental and theoretical studies
of Lorentz and CPT violation
involving 
mesons,\cite{hadronexpt,kpo,hadronth,ak}
baryons,\cite{ccexpt,spaceexpt,cane}
electrons,\cite{eexpt,eexpt2,eexpt3}
photons,\cite{photon,km02}
muons,\cite{muons}
and the Higgs sector.\cite{higgs}
We remark 
that neutrino-oscillation experiments
offer the potential for discovery.\cite{sme,neutrinos,nulong}
A number of these studies 
involve some form of antimatter.
CPT and Lorentz tests with antimatter 
will be discussed further in Sec.\ \ref{test}.

\section{Sample mechanisms for Lorentz and CPT violation}
\label{mech}

In the previous two sections,
we have studied various general {\it types} of manifestations 
of Lorentz and CPT breakdown,
as well as the {\it description} 
of the corresponding effects in a microscopic model,
such as the SME.
However,
the question of {\it how} exactly a Lorentz- and CPT-invariant candidate theory
can lead to the violation of these symmetries
has thus far been left unaddressed.
The purpose of this section
is to provide some intuition
about such mechanisms for Lorentz and CPT breaking in underlying physics.
Of the various possible mechanisms
mentioned in Sec.\ \ref{mot},
we will focus on spontaneous Lorentz violation
and Lorentz breakdown through varying scalars.

\begin{figure}
\begin{center}
\includegraphics[width=0.95\hsize]{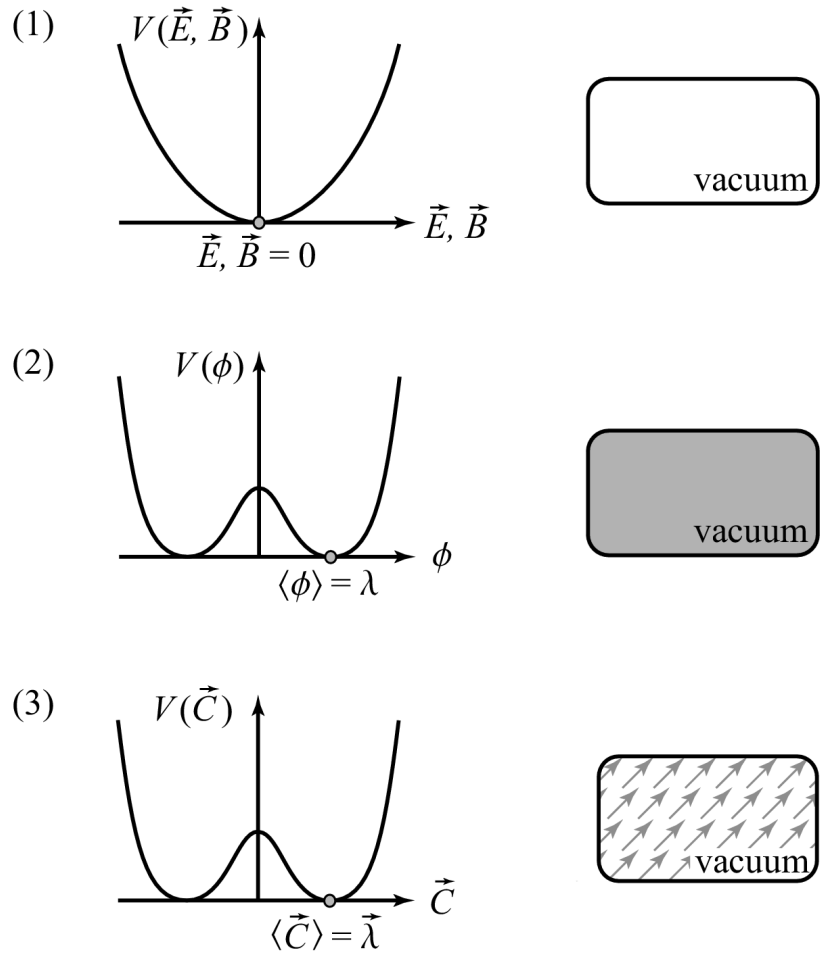}
\end{center}
\caption{Spontaneous symmetry breaking.
In conventional electrodynamics (1),
the lowest-energy state is attained 
for zero $\vec{E}$ and $\vec{B}$ fields.
The vacuum remains essentially empty.
For the Higgs-type field (2),
interactions lead to an energy density $V(\phi)$
that forces a nonzero value of $\phi$ in the ground state.
The vacuum fills with a scalar condensate shown in gray.
Lorentz invariance still holds 
(other, internal symmetries may be violated though).
Vector fields occurring, e.g., in string theory (3)
can have interactions similar to those of the Higgs
requiring a nonzero field value in the lowest-energy state. 
The VEV of a vector field selects a preferred direction in the vacuum, 
which has now properties paralleling those of a crystal.
}
\label{fig3} 
\end{figure}

{\bf Spontaneous Lorentz and CPT violation.}
The mechanism of spontaneous symmetry breaking
is well established in various subfields of physics,
such as the physics of elastic media, 
condensed-matter physics, 
and elementary particle theory.
From a theoretical perspective,
this mechanism is very attractive
because the invariance is essentially violated 
through a non-trivial ground-state solution.
The underlying dynamics of the system
governed by the Hamiltonian
remains completely invariant under the symmetry.
To gain intuition 
about spontaneous Lorentz and CPT breakdown,
we will consider three sample systems,
whose features will gradually lead us 
to a better understanding of the effect.
An illustration
supporting these three examples
is given in Fig.\ \ref{fig3}.

First, let us consider classical electrodynamics.
Any electromagnetic-field configuration
is associated with an energy density $V(\vec{E},\vec{B})$ 
given by
\beq
\label{max_en_den}
V(\vec{E},\vec{B})=\fr{1}{2} \left(\vec{E}^2+\vec{B}^2\right)\, ,
\eeq
where we have employed natural units,
and $\vec{E}$ and $\vec{B}$
denote the electric and magnetic field,
respectively.
With Eq.\ \rf{max_en_den},
we can determine the field energy 
of any given solution of the Maxwell equations. 
Note 
that if the electric field, or the magnetic field, or both
are nonzero somewhere in spacetime,
the energy stored in these fields will be strictly positive.
The field energy can only be exactly zero
when both $\vec{E}$ and $\vec{B}$ vanish everywhere.
The vacuum (or ground state)
is usually identified with the lowest-energy configuration of a system.
We see that in conventional electrodynamics
the configuration with the lowest energy
is the field-free one,
so that the Maxwell vacuum is empty
(disregarding quantum fluctuations).

Second,
let us look at the Higgs field,
which is part of the phenomenologically very successful 
Standard Model of particle physics.
Unlike the electromagnetic field,
the Higgs field is a scalar.
In what follows,
we can adopt some simplifications 
without distorting 
the features important in the present context.
The expression for the energy density of our Higgs scalar $\phi$
in situations with spacetime independence 
is given by 
\beq
\label{scal_en_den}
V(\phi)=(\phi^2-\lambda^2)^2\, ,
\eeq
where $\lambda$ is a constant.
As in the Maxwell case discussed above,
the lowest possible field energy is zero.
Note, however,
that this configuration {\it requires} 
$\phi$ to be nonzero: $\phi=\pm\lambda$.
It follows 
that the vacuum for a system containing a Higgs-type field
is not empty;
it is, in fact, 
filled with constant scalar field 
$\phi_{vac}\equiv\langle\phi\rangle=\pm\lambda$.
In quantum theory,
the quantity $\langle\phi\rangle$
is called the vacuum expectation value (VEV)
of $\phi$.
One of the physical effects 
caused by the VEV of the Higgs
is to give masses to many elementary particles.
Note, however,
that $\langle\phi\rangle$ is a scalar
and does {\it not} select a preferred direction in spacetime.

Third, we consider a vector field $\vec{C}$
(the relativistic generalization is straightforward)
not contained in the Standard-Model.
Of course, 
there is no observational evidence for such a field
at the present time,
but fields like $\vec{C}$
frequently arise in approaches to more fundamental physics.
In analogy to Higgs case,
we take its expression for energy density 
in situations with constant $\vec{C}$
to be
\beq
\label{vec_en_den}
V(\vec{C})=(\vec{C}^2-\lambda^2)^2\, .
\eeq
As in the previous two examples,
the lowest possible energy is exactly zero.
As for the Higgs,
this lowest energy configuration is attained for nonzero $\vec{C}$.
More specifically, 
we must have $\vec{C}_{vac}\equiv\langle\vec{C}\rangle=\vec{\lambda}$,
where $\vec{\lambda}$ is any constant vector satisfying $\vec{\lambda}^2=\lambda^2$.
Again,
the vacuum is not empty,
but filled with the VEV of our vector field.
Since we have only considered 
constant solutions $\vec{C}$,
$\langle\vec{C}\rangle$
is also spacetime independent 
($x$ dependence would lead to 
positive definite derivative terms 
in Eq.\ \rf{vec_en_den} raising the energy density).
The true vacuum in our model
therefore contains an intrinsic direction
determined by $\langle\vec{C}\rangle$
{\it violating rotation invariance and thus Lorentz symmetry}.
We mention 
that interactions leading to energy densities like \rf{vec_en_den}
are absent in conventional renormalizable gauge theories,
but can be found in the context of strings, for example.

{\bf Cosmologically varying scalars.}
A spacetime-dependent scalar,
regardless of the mechanism driving the variation,
typically implies the breaking of spacetime-translation invariance.
Since translations and Lorentz transformations
are closely linked in the Poincar\'e group,
it is reasonable to expect
that the translation-symmetry violation
also affects Lorentz invariance.

Consider,
for instance,
the angular-momentum tensor $J^{\mu\nu}$,
which is the generator of Lorentz transformations:
\beq
J^{\mu\nu}=\int d^3x \;\big(\th^{0\mu}x^{\nu}-\th^{0\nu}x^{\mu}\big).
\label{gen}
\eeq
Note
that this definition 
contains the energy--momentum tensor $\th^{\mu\nu}$,
which is not conserved
when translation invariance is broken.
In general,
$J^{\mu\nu}$
will possess a nontrivial dependence on time,
so that the usual time-independent 
Lorentz-transformation generators do not exist.
As a result,
Lorentz and CPT symmetry
are no longer assured.

\begin{figure}
\begin{center}
\includegraphics[width=0.95\hsize]{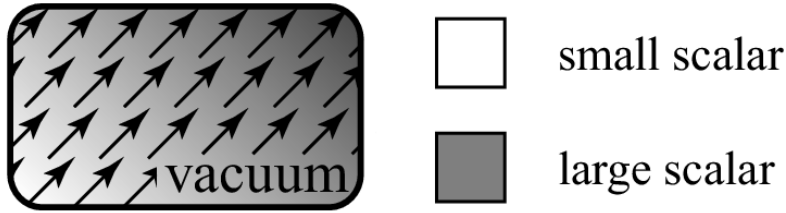}
\end{center}
\caption{Lorentz breakdown through varying scalars.
The value of the scalar corresponds to the background shade of gray: 
the darker regions are associated with greater values of the scalar.
The gradient represented by the black arrows 
selects a preferred direction in the vacuum.
Although Lorentz coordinate independence is maintained,
particle Lorentz invariance is violated.
}
\label{fig4} 
\end{figure} 

Intuitively,
the violation of Lorentz invariance 
in the presence of a varying scalar can be understood as follows.
The 4-gradient of the scalar must be nonzero
in some regions of spacetime. 
Such a gradient 
then selects a preferred direction in this region (see Fig.\ \ref{fig4}).
Consider, 
for example, 
a particle
that interacts with the scalar.
Its propagation features
might be different
in the directions parallel and perpendicular to the gradient.
Physically inequivalent directions
imply the violation of rotation symmetry.
Since rotations are contained in the Lorentz group,
Lorentz invariance must be violated.

Lorentz violation induced by varying scalars
can also be established at the Lagrangian level.
Consider, 
for instance,
a system with a varying coupling $\xi(x)$
and scalar fields $\phi$ and $\Phi$,
such that the Lagrangian $\mathcal{L}$ contains a term
$\xi(x)\,\partial^{\mu}\phi\,\partial_{\mu}\Phi$.
The action for this system can be integrated by parts
(e.g., with respect to the first partial derivative in the above term)
without affecting the equations of motion.
An equivalent Lagrangian $\mathcal{L}'$ would then obey
\begin{equation}
\mathcal{L}'\supset -K^{\mu}\phi\,\partial_{\mu}\Phi\, ,
\label{example1}
\end{equation}
where $K^{\mu}\equiv\partial^{\mu}\xi$ is an external
nondynamical 4-vector,
which clearly violates Lorentz symmetry.
We remark
that for variations of $\xi$ on cosmological scales,
$K^{\mu}$ is constant to an excellent approximation 
locally---say on solar-system scales.

\section{Lorentz and CPT tests involving antimatter}
\label{test}

\begin{figure}
\begin{center}
\includegraphics[width=0.95\hsize]{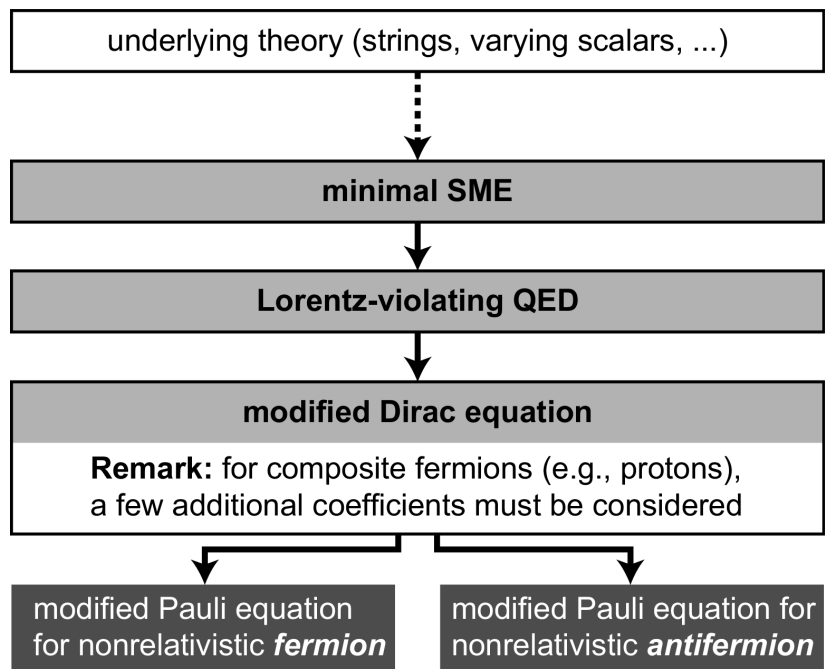}
\end{center}
\caption{SME analysis of atomic spectra.
The resulting modified Pauli equation
for fermions and that for antifermions---which typically 
differ---are employed to describe the (anti)proton
and the orbiting (anti)electron in the H ($\bar{\textrm{H}}$) system.
}
\label{fig5} 
\end{figure}

Numerous Lorentz and CPT tests 
among those listed in Sec.\ \ref{smesec}
involve some form of antimatter.
As pointed out earlier,
certain matter--antimatter comparisons
are extremely sensitive to CPT violations. 
This is unsurprising
because CPT symmetry connects particles and antiparticles.
CPT tests 
with subatomic particles
typically involve quantum numbers like mass, charge, spin, etc.
Atoms and their anti-atoms 
possess additional, qualitatively different properties,
such as spectra, 
that can be compared.
The possibility of $\bar{\textrm{H}}$ production
combined with the ultrahigh sensitivities attainable in atomic spectroscopy 
and the simplicity of the two-body problem 
(antiproton nucleus and orbiting positron)
make this anti-atom particularly well suited 
for such investigations.
The determination of SME predictions 
for such physical systems
follows the outline shown in Fig.\ \ref{fig5}.

{\bf The unmixed 1S--2S transition.}
The experimental resolution of the transition
involving the unmixed spin states
is expected to be about one part in $10^{-18}$.
This sensitivity appears promising 
in light of potential Planck-suppressed quantum-gravity effects.
On the other hand,
the leading-order SME calculation
shows identical shifts
for free H or $\overline{\textrm{H}}$
in the initial and final states
with respect to the conventional levels.
It follows 
that this transition is less suitable
for the measurement of unsuppressed Lorentz- and CPT-breaking effects.
The largest non-trivial contribution 
to this transition
within the SME test framework
arises through relativistic corrections,
and it involves two additional powers 
of the fine-structure parameter $\alpha=\frac{1}{137}$.
The expected energy shift---already at zeroth order in $\alpha$
anticipated to be minute---comes therefore 
with an additional suppression 
by a factor of more than ten thousand.
This further exemplifies the need and importance
of a viable test model for Lorentz- and CPT-violation searches.

\begin{figure}[h]
\begin{center}
\includegraphics[width=0.95\hsize]{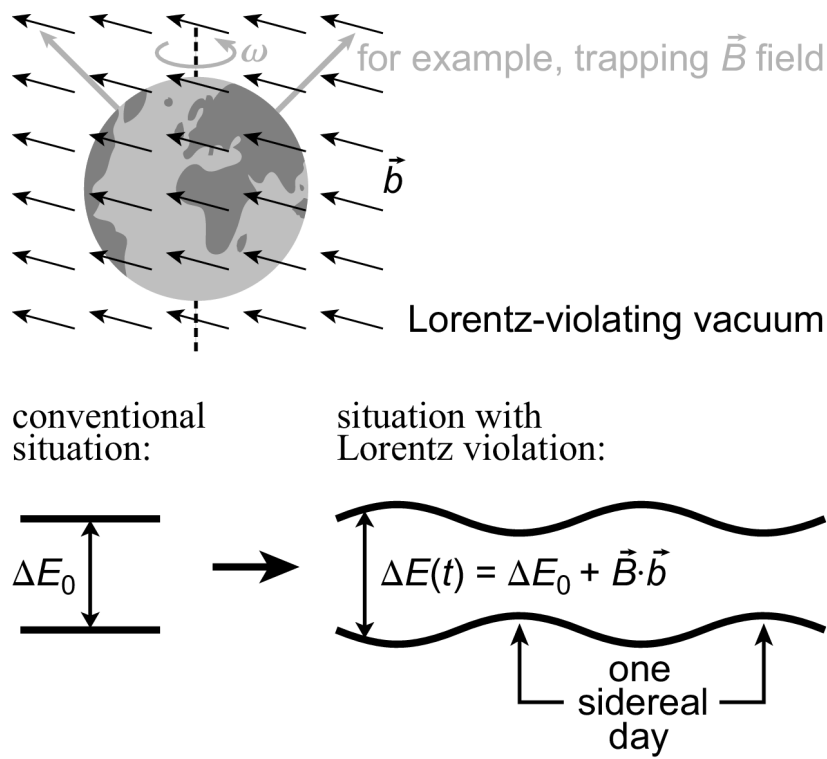}
\end{center}
\caption{Sidereal variations. 
Experiments are typically associated with an intrinsic direction.
For instance, particle traps usually involve a magnetic field.
As the Earth rotates, this direction will change
if the experiment is attached to the Earth. 
In the above figure,
a trapping field $\vec{B}$ pointing vertically upward is shown 
at two times separated by approximately 12 hours (gray arrows).
The angle between the Lorentz-violating background
(black $\vec{b}$ arrows) and the orientation of the experiment
is clearly different at these two times.
An observable, such as an atomic transition,
may for example acquire a correction $\sim\vec{B}\cdot\vec{b}$
that leads to the shown sidereal modulation.
}
\label{fig6} 
\end{figure} 

{\bf The spin-mixed 1S--2S transitions.}
When H or $\overline{\textrm{H}}$ 
is confined with magnetic fields---such as in a Ioffe--Pritchard
trap---the 1S and the 2S levels are each split 
by the Zeeman effect.
In the framework of the SME,
one can show 
that in this situation the 1S--2S transition
between the spin-mixed states
is affected 
by Lorentz and CPT violation
at leading order.
A disadvantage from a practical perspective is
the field dependence of this transition,
so that the experimental resolution
is limited 
by the size of the inhomogeneity 
in the trapping magnetic field.
The development of novel experimental techniques 
would appear necessary
to achieve resolutions 
close to the natural linewidth.

{\bf Hyperfine Zeeman transitions within the 1S state.} 
An alternative Lorentz and CPT test
could measure the transition frequency 
between the Zeeman-split states
within the 1S level itself.
Even in the zero-magnetic-field limit,
the SME predicts first-order effects
for two of these transitions.
Other transitions of this type,
such as the conventional Hydrogen-maser line,
can be well resolved in experiments.

{\bf Tests in Penning traps.}
The SME predicts
that not only atomic energy levels
can be shifted in the presence of Lorentz violation,
but also proton and antiproton levels
in Penning traps.
A calculation shows
that only one SME coefficient
($b^{\mu}$ in the standard notation)
leads to transition-frequency differences
between the proton and its antiparticle.
More specifically,
the anomaly frequencies
are changed in opposite directions
for protons and antiprotons.
This effect permits a clean observational bound
on $b^{\mu}$ for the proton.

{\bf Searches for sidereal variations.}
Another general signature for Lorentz and CPT breakdown
is the variation of measured quantities
with the sidereal day.
The anti-CPT theorem implies that CPT breakdown
always comes with Lorentz violation,
which in turn is typically accompanied by the loss of isotropy.
Thus, experimental effects will generally depend on the direction.
As the laboratory is attached to the rotating Earth,
its orientation will change continually
leading to sidereal modulations of signals.
This situation is schematically depicted in Fig.\ \ref{fig6}.
Note that sidereal-variation tests 
are not confined exclusively to H--$\bar{\textrm{H}}$ spectroscopy,
but they can also be performed 
in the context of other rotation-violation searches.
Recent experiments with H-masers 
employing ingenious experimental techniques
are based on such modulations.\cite{hum}

\section{Summary}
\label{summ}

Although Lorentz and CPT symmetry 
are deeply ingrained 
in the currently accepted laws of physics, 
there are a variety of candidate underlying theories 
that could generate the breakdown of these symmetries. 
The sensitivity attainable in matter--antimatter comparisons
offers the possibility for CPT-violation searches
with Planck precision.
Lorentz tests open an additional avenue
for CPT measurements
because CPT breakdown implies Lorentz violation.

A potential source of Lorentz and CPT violation
is spontaneous symmetry breaking in string theory.
Since this mechanism is theoretically very attractive
and since strings show great potential as a candidate fundamental theory,
this Lorentz-violation source is particularly promising.
Lorentz and CPT violation
can also originate from
spacetime-dependent couplings:
the gradient of such couplings
selects a preferred direction 
in the effective vacuum. 
This mechanism for Lorentz breaking
might be of interest
in light of recent claims of a time-dependent fine-structure parameter
and the presence of varying scalar fields
in many cosmological models.

The leading-order Lorentz- and CPT-violating effects 
resulting from Lorentz-symmetry breakdown 
in approaches to fundamental physics
are described by the SME. 
At the level of effective quantum field theory, 
the SME 
is the most general dynamical framework 
for Lorentz and CPT breaking 
that is compatible 
with the fundamental principle of coordinate independence. 
Experimental investigations
are therefore best performed within the SME. 

Cold antiprotons
are excellent high-sensitivity tools
in experimental searches for Planck-scale physics.
Suppressed and unsuppressed effects exist 
for 1S--2S transitions in H and $\overline{\rm H}$.
Leading-order shifts
are also predicted in the 1S hyperfine Zeeman levels,
which offers the possibility of alternative measurements.
Further possibilities for Lorentz- and CPT-violation searches
with antiprotons
exist in Penning traps,
where anomaly frequencies are affected.
In general,
tests with cold antiprotons probe parameter combinations
inaccessible by other experiments.

\section*{Acknowledgments}
The author would like to thank Yasunori Yamazaki
for organizing this stimulating meeting,
for the invitation to attend,
and for partial financial support.

\end{document}